# $Fe_2O_3$/ $Co_3O_4$ composite nanoparticle ethanol sensor

**Ali Mirzaei[1], Sunghoon Park[2], Gun-Joo Sun[2], Hyejoon Kheel[2], Sangmin Lee[3], Chongmu Lee[*2]**

*[1]Department of Materials Science and Engineering, Shiraz University, Shiraz, Iran*

*[2]Department of Materials Science and Engineering, Inha University, 253 Yonghyun-dong, Nam-gu, Incheon 402-751, Republic of Korea*

*[3]Department of Electronic Engineering, Inha University, 253 Yonghyun-dong, Nam-gu, Incheon 402-751, Republic of Korea*

* Corresponding Author: cmlee@inha.ac.kr

In this study $Fe_2O_3$/$Co_3O_4$ nanocomposites were synthesized by using a simple hydrothermal route. The X-ray diffraction analysis results showed that synthesized powders were pure, and nanocrystalline in nature. Moreover, Scanning electron microscopy revealed that $Fe_2O_3$ nanoparticles had spherical shapes while $Co_3O_4$ particles had a rod-like morphology. Ethanol sensing properties of $Fe_2O_3$/$Co_3O_4$ nanocomposites were examined and compared with those of pristine $Fe_2O_3$ nanoparticles. It was shown that the gas sensing properties of $Fe_2O_3$/$Co_3O_4$ nanocomposites were superior to those of pristine $Fe_2O_3$ nanoparticles and for all concentrations of ethanol, the response of the nanocomposite sensor was higher than the pristine $Fe_2O_3$ nanoparticle sensor. In detail, the response of $Fe_2O_3$/$Co_3O_4$ nanocomposite sensor to 200 ppm of ethanol at 300°C was about 3 times higher than pristine one. Also in general, the response and recovery times of $Fe_2O_3$/$Co_3O_4$ nanocomposite sensor were shorter than those of the pristine one. The improved sensing characteristics of the $Fe_2O_3$/$Co_3O_4$ sensor were attributed to a combination of several effects: the formation of a potential barrier at the $Fe_2O_3$-$Co_3O_4$ interface, the enhanced modulation of the conduction channel width accompanying the adsorption and desorption of ethanol gas, the catalytic activity of $Co_3O_4$ for the oxidation of ethanol, the stronger oxygen adsorption of p-type $Co_3O_4$, and the formation of preferential adsorption sites.

PACS number: 81.07.-b, 81.05.Ea, 81.15.Gh, 85.35.-p

Keywords: Gas sensor, $Fe_2O_3$, $Co_3O_4$, nanoparticles, nanocomposite, mixed oxide, gas sensor.

## I. INTRODUCTION

Because of the peculiar properties originated from their individual phases, mixed metal oxide composites are of special interest and they have recently emerged as promising candidates for gas detection [1]. It has been realized that such systems may benefit from a combination of the best sensing properties of the pure components. A combination of more than two different kinds of oxides leads to the modification of the electronic structure of the system. This includes the changes in the bulk as well as in the surface properties. Surface properties are expected to be influenced by new interfaces of the two different oxides with different chemical compositions in contact. It is anticipated that these phenomena will contribute advantageously to the gas sensing mechanism [2].

Among the heterostructured semiconducting metal oxides for gas sensing application, $Fe_2O_3/Co_3O_4$ system is very interesting. $Fe_2O_3$ (hematite) is one of the cheapest semiconducting materials (n-type, $E_g$ = 2.1 eV) and thanks to its low cost, high resistance to corrosion, and nontoxicity properties, this most stable iron oxide has been traditionally used as catalysts, pigments, electrode materials and gas sensor [3-6]. On the other hand, $Co_3O_4$ with spinel structure is an important p-type semiconductor with an indirect band gap of 1.5 eV [7] and it is well-known as a good catalyst in the oxidation reaction of CO [8]. Although it is known that a good catalyst does not have to be a good sensor, $Co_3O_4$ has attracted some research interest in the last decade as a candidate of a gas sensor [9-11].

Ethanol ($C_2H_5OH$), an inflammable chemical compound, is one of the most commonly and widely used alcohols and has many applications in food, biomedical, transportation and chemical industries as well as health and safety [12]. Maximum recommended exposure level of ethanol according to the Occupational Safety Health and Administration (OSHA) established to be 1000 ppm [13]. Exposure to ethanol vapor results in health problems such as difficulty in breathing, headache, drowsiness, irritation of eyes, liver damage and so forth [14]. Accordingly demand for ethanol detection is very high.

This paper deals with the synthesis, characterization and ethanol sensing performance of pristine $Fe_2O_3$ and $Fe_2O_3/Co_3O_4$ composite nanoparticles. The $Fe_2O_3$ and $Fe_2O_3/Co_3O_4$ composite nanoparticles were synthesized via a facile hydrothermal route and were used for the sensitive detection of ethanol. The $Fe_2O_3/Co_3O_4$ composite nanoparticle sensor showed enhanced sensing performance in terms of the response and response/recovery times compared to the pristine $Fe_2O_3$ nanoparticle sensor.

## II. EXPERIMENTAL PROCEDURE

### 1. Synthesis of $Fe_2O_3$ and $Co_3O_4$ nanoparticles.

To prepare 50-mM iron (III) chloride ($FeCl_3$) solution, iron (III) chloride was dissolved in deionized water and stirred for 2 hr at 50°C. In a separate flask NaOH was dissolved in deionized water to prepare 50 ml of 50 mM NaOH solution. Then, two solutions were mixed and this solution was poured into a Chadorok autoclave and maintained at 160°C for 15 h. afterwards, the solution was removed using a solution aspirator, leaving a brown powder behind. The synthesized powders were washed with deionized water: acetone: isopropanol alcohol = 1: 1: 1 mixture. The synthesized $Fe_2O_3$ nanopowders were dried in an oven at 120°C for 12 h and heat treated in a vacuum furnace (1mTorr) at 500°C for 1 h. In a similar manner, $Co_3O_4$ nanoparticles were prepared using cobalt acetate ($Co(C_2H_3O_2)_2(H_2O)_4$) and NaOH.

### 2. Fabrication of pristine $Fe_2O_3$ and $Fe_2O_3/Co_3O_4$ composite nanoparticle sensors

40 mg of $Fe_2O_3$ nanoparticles and 10 mg $Co_3O_4$ nanoparticles were dispersed in 50 ml isopropyl alcohol and ultrasonicated for 1 h. 1 mL drop of the solution containing $Fe_2O_3$ and $Co_3O_4$ nanoparticles were placed onto an interdigital electrode (IDE) pattern (size: 10 mm × 10 mm) and dried at 150°C in air for 1 h.

### 3. Materials characterization

The crystallinity and phases of the pure $Fe_2O_3$ nanoparticles and $Fe_2O_3/Co_3O_4$ composite nanoparticles were analyzed by X-ray diffraction (XRD, Philips X'pert MRD) using Cu K$\alpha_1$ radiation (1.5406Å), and the morphology and particle size of the samples were examined by scanning electron microscopy (SEM, Hitachi S-4200).

### 4. Sensing tests

During the measurements, the gas sensors were placed in a gas chamber with an electrical feed through. A pre-determined amount of ethanol vapor was injected into the gas chamber to obtain ethanol concentrations of 5, 10, 20, 50, 100, and 200 ppm while simultaneously the electrical resistance of the sensor was monitored. The response was defined as $R_a/R_g$, where $R_a$ and $R_g$ are the electrical resistances of sensor in air and ethanol, respectively. The response and recovery times were defined as the times to reach 90 % of the resistance change upon exposure to ethanol and air, respectively.

## III. Results and Discussions

### 1. SEM studies

The SEM images of pure $Co_3O_4$ nanoparticles, pure $Fe_2O_3$ nanoparticles and $Fe_2O_3/Co_3O_4$ nanocomposite are presented in Figs. 1(a), (b) and (c), respectively. The $Fe_2O_3$ nanoparticles have a spherical shape, whereas the $Co_3O_4$ nanoparticles have elongated rod or plate-like shape. The image of the composite nanoparticles shows almost spherical-shaped $Fe_2O_3$ nanoparticles with diameters ranging from 20 nm to 70 nm, whereas $Co_3O_4$ particles are elongated in two different directions as plates with widths of 20-80 nm and lengths of 0.1-0.4 µm. Furthermore, it is observed that individual $Fe_3O_4$ and $Co_3O_4$ particles are mixed relatively intimately, i.e., they are contact each other. As will be discussed later, intimate mixing could improve gas sensing properties of the fabricated nanocomposite sensor. Figures 1(d) and (e) show EDX elemental mapping and EDX spectrum of $Fe_2O_3$/ $Co_3O_4$ nanocomposite, respectively. The Fe map in Fig. 1(b) reveals that $Fe_2O_3$ particles are distributed relatively uniformly, whereas the Co map in Fig. 1(b) shows that $Co_3O_4$ nanoparticles have an elongated hemisphere shape. The EDX spectrum (Fig. 1(c)) indicates that the composite nanoparticles are composed of Co, Fe and O. the Cu and Cr peaks in the spectrum are due to the Cu/Cr electrode.

**2. XRD studies**

X-ray diffraction (XRD) measurements were made to determine the crystal structure of the synthesized powders. Figure 2 shows the XRD patterns of $Fe_2O_3$ nanoparticles and $Fe_2O_3/Co_3O_4$ nanocomposites. The XRD pattern of $Fe_2O_3$ exhibits six diffraction peaks at $2\theta = 33.106°$ (104), $2\theta = 35.559°\ (110)$, $2\theta = 50.634°\ (12\bar{4})$, $2\theta = 49.428°$ (113), $2\theta = 62.385°\ (214)$, $2\theta = 65.981°\ (300)$, assigned to the rhombohedral-structured $Fe_2O_3$ (JCPDS No: 89- 2810).

In the XRD pattern of the nanocomposite, there are three additional peaks. The measured 2θ values of 36.458°, 52.628° and 65.728° coincide well with the reference data of $Co_3O_4$ crystals. However, there are no observed peaks at 49.65° and 72.958° correspond to CoO crystals. This shows that cobalt oxide exist as a phase of $Co_3O_4$ in the final nanocomposite.

According to above analysis, all the diffraction peaks can be indexed to the lattice planes of $Fe_2O_3$ and $Co_3O_4$, suggesting that the synthesized nanostructures had a high purity without containing other compounds or impurities.

The XRD patterns in Fig. 2 showed broad peaks, indicating the existence of nanocrystals. To determine the crystallinity of the synthesized nanostructures, the crystallite size was estimated using the Scherrer formula:

$$D = 0.9\left(\frac{\lambda}{\beta\cos\theta}\right) \quad (1)$$

,where D is the crystallite size in nm, λ is the wavelength of X-rays used (1.5406Å), β is the full-width at half maximum in degree and θ is the diffraction angle in degree. The (110) plane of $Fe_2O_3$ nanoparticles was chosen to calculate the crystallite size and the calculated crystallite size was 60 nm and ~ 75 nm for $Fe_2O_3$ nanoparticles and $Fe_2O_3/Co_3O_4$ nanocomposites, respectively.

**3. TEM Analysis**

Figures 3 (a), (b) and (c) show the low-magnification TEM image of $Fe_2O_3$/ $Co_3O_4$ nanocomposites, the High Resolution-TEM image of the interfacial region of $Fe_2O_3$/ $Co_3O_4$ nanocomposites and the corresponding selected area diffraction pattern, respectively. The $Fe_2O_3$ nanoparticles are smaller and spherical, whereas the $Co_3O_4$ nanoparticles are larger and elongated (Fig. 3(a)). The fringes in the HRTEM image (Fig. 3(b)) and the concentric ring patterns reveal that both the $Fe_2O_3$ and $Co_3O_4$ nanoparticles in the composites are polycrystalline.

**3. Sensing studies**

**3.1 Optimal working temperature**

From the application point of view, one wishes to minimize the power consumption needed for the sensor operation. Therefore, it is important to determine the optimal working temperature of a sensor. Figure 4 shows the temperature dependence of the response of the two sensors to 200 ppm of ethanol gas at temperatures ranging from 250°C to 350°C. For all temperatures, the response of the $Fe_2O_3/Co_3O_4$ nanocomposite sensor was higher than that of the pristine $Fe_2O_3$ sensor. For both sensors the response increased with increasing the operating temperature up to 300°C and then decreased. The sensor response to ethanol gas depends on a delicate balance between the adsorption and desorption rates of ethanol and the surface reactivity of adsorbed ethanol with adsorbed oxygen species. The increase in the operating temperature facilitates ethanol adsorption to a certain extent, and the reaction rate occurring on the sensor surface leads to enhanced gas response. At higher temperatures, the gas response decreases due to the desorption of ethanol, which decreases the amount of ethanol adsorbed on the sensor surface. After determining the optimal temperatures, all the gas sensing tests were performed at 300°C for both sensors.

**3.2 Sensor response with ethanol gas concentration**

After determining the optimum working temperature of the two sensors, the responses of both sensors to different concentrations of ethanol (5, 10, 20, 50, 100 and 200 ppm) at 300°C was investigated. Figures 5(a) and (b) show the gas response transients of the pristine $Fe_2O_3$ nanoparticle and $Fe_2O_3/Co_3O_4$ nanocomposite sensors to different concentrations of ethanol gas at 300°C. The nanocomposite sensor showed higher resistance than the pristine $Fe_2O_3$ nanoparticle sensor, which might be due to the higher intrinsic resistance of $Co_3O_4$ than $Fe_3O_4$. Exposure of both sensors to ethanol gas led to a decrease in resistance, suggesting that the addition of $Co_3O_4$ nanoparticles to the $Fe_2O_3$ nanoparticles does not change the n-type semiconducting properties of the pristine $Fe_2O_3$ nanoparticles. It is also observed that the signal returns to its initial baseline value after each pulse. This observation indicates that the adsorption of ethanol on the surface layer is fully reversible.

Figure 6(a) presents the calibration curve of both sensors at 300°C. For all concentrations of ethanol gas, the $Fe_2O_3$/ $Co_3O_4$ nanocomposite sensor was more sensitive to ethanol gas than the pristine $Fe_2O_3$ sensor. As shown at low concentrations of ethanol gas, the difference between the responses of the two sensors is negligible, but at higher concentrations, the response of the $Fe_2O_3/Co_3O_4$ nanocomposite sensor was far higher than that of the pristine $Fe_2O_3$ sensor. The relationship between the sensor response ($S=R_a/R_g$) and ethanol concentration ($C_{ethanol}$) can be written as:

$$S = A[C_{ethanol}]^b + 1 \tag{2}$$

where A, b, and $[C_{ethanol}]$ are a constant, an exponent and the ethanol concentration, respectively.

According to the above formula, the response is directly proportional to the ethanol concentration. Figure 6(b) shows a logarithmic plot of the data in Fig. 6(a). Plot of ln (s-1) versus ln (c) for $Fe_2O_3/Co_3O_4$ nanocomposite sensor gives an almost a straight line. However this plot for $Fe_2O_3$ sensor does not give a straight line, demonstrating the good behavior of the nanocomposite sensor according to the theory of power laws for semiconductor sensors [15].

### 3.4 Sensing Mechanism

When $Fe_2O_3$ and $Fe_2O_3/Co_3O_4$ nanocomposite sensors are exposed to the air, oxygen molecules are adsorbed on the surface and extract electrons from the conduction band and the electron depletion region extends from the surface, which increases the resistance of sensors. The reaction kinetics may be explained by the following reactions [16]:

$$O_{2(gas)} \rightarrow O_{2(ads)} \quad (3)$$

$$O_{2(ads)} + \bar{e} = O^{-}_{2(ads)} \quad (4)$$

$$O^{-}_{2(ads)} + \bar{e} = 2O^{-}_{(ads)} \quad (5)$$

$$O^{-}_{(ads)} + \bar{e} = O^{2-}_{(ads)} \quad (6)$$

After exposing the sensors to ethanol vapor, the ethanol molecules could be adsorbed on the surfaces of sensors and react with the adsorbed oxygen species to form water vapor and $CO_2$ (Eqs. (7)-(9)). This leads to an increase in concentration of electrons. This eventually decreases the resistivity of the sensor which can be used for the detection of ethanol gas [17].

$$C_2H_5OH_{(ads)} + O^{-}_{(ads)} = CH_3CHO_{(ads)} + H_2O + \bar{e} \quad (7)$$

$$CH_3CHO_{(ads)} + 5O^{-}_{(ads)} = 2CO_2 + 2H_2O + 5\bar{e} \quad (8)$$

$$CH_3CHO_{(ads)} + 6O^{2-}_{(ads)} = 2CO_2 + 3H_2O + 12\bar{e} \quad (9)$$

The gas sensing properties of the $Fe_2O_3/Co_3O_4$ nanocomposite sensor towards ethanol were superior to those of the pristine $Fe_2O_3$ sensor. These enhanced ethanol sensing properties might be due mainly to a combination of the following effects: (i) stronger adsorption of oxygen molecules by p-type $Co_3O_4$: p-type metal oxide semiconductors only chemisorb as much oxygen as possible to compensate for their deficiencies. On the other hand, the concentration of surface oxygen on p-type semiconductors is significantly higher than that of n-type semiconductors [10] (ii) creation of preferential adsorption sites for oxygen and ethanol molecules: crystallographic defects are created at the $Fe_2O_3/Co_3O_4$ interface due to the lattice mismatch between the two materials, which provides preferential adsorption sites for oxygen and ethanol molecules [18] (iii) enhanced modulation of the depletion layer width accompanying the adsorption and desorption of ethanol gas and [19,20] and (iv) It is well known that the combination of a p-type semiconductor (such as $Co_3O_4$) with an n-type semiconducting oxide (such as $Fe_2O_3$) can form a p-n junction. For gas sensing applications, the effective integration of p- and n-type semiconductors can provide higher sensing responses because of the formation of a deeper extended depletion. Therefore, the large modulation of the potential barrier height at the $Fe_2O_3/Co_3O_4$ interface (p-n junction) accompanying the adsorption and desorption of ethanol vapor.

Modulation of the conduction channel width occurs accompanying the adsorption and desorption of ethanol. The change in depletion layer width is slightly larger in the composite nanoparticle sensor ($W_{D4}$-$W_{D3}$) than in its pure $Fe_2O_3$ counterpart ($W_{D2}$-$W_{D1}$), as shown in Figs. 6(a) and (b). Also a potential barrier forms at the $Fe_2O_3/Co_3O_4$ p-n junction and potential barrier height modulation occurs during the adsorption and desorption of acetone gas. The differences in the potential barrier heights at the n-n and p-n junctions between in air and in acetone gas are $V_2$-$V_1$ and $V_4$-$V_3$, respectively, and the latter is slightly larger than the former ($V_4$-$V_3$ > $V_2$-$V_1$), as shown in Fig. 6. The resistance of the sensor is related to the potential barrier height using the following equation:

$$R = R_0 \exp(qV/kT) \quad (10)$$

,where R is the resistance of the material, $R_0$ is the baseline resistance, q is the charge of an electron, V is the potential energy barrier height, k is Boltzmann’s constant, and T is the absolute temperature of the sensing material. Because the response is determined by $R_a/R_g$, the response depends on the potential barrier height at the $Fe_2O_3/Co_3O_4$ interface. Furthermore, the oxidative catalytic activity of $Co_3O_4$ is well-known. $Co_3O_4$ expedites the oxidation reaction of ethanol leading to an enhanced response to ethanol. Finally crystallographic defects are created at the $Fe_2O_3/Co_3O_4$ interface due to the lattice mismatch between the two materials, which provides preferential adsorption sites for oxygen and ethanol molecules [10].

### 3.5 Response and Recovery times

Response and recovery times of $Fe_2O_3$ nanoparticle sensors and $Fe_2O_3/Co_3O_4$ nanocomposites sensors are shown in Fig 8(a) and (b) respectively. For both recovery and response times both sensors show short times indicating fast capabilities of ethanol detection which is a required necessity for practical applications. In the case of response time, in general, the $Fe_2O_3/Co_3O_4$ nanocomposite sensor has a shorter response time, which is probably due to the higher resistance of the $Fe_2O_3/Co_3O_4$ nanocomposite sensor at 300°C, which means that there are more adsorbed oxygen species on the surface of nanocomposite sensor, so that after injection of ethanol gas, they react very fact with ethanol and consequently response time becomes very short. The recovery time of the $Fe_2O_3/Co_3O_4$ nanocomposite sensor is longer than that of the pristine $Fe_2O_3$ sensor. The recovery reaction consists of the diffusion of oxygen gas to the sensing surface, the adsorption of oxygen molecules, the dissociation of oxygen molecule into atomic oxygen, and the ionization of atomic oxygen. The slower recovery time can be explained by the sluggish surface reactions regarding the adsorption, dissociation, and ionization of oxygen at $Fe_2O_3/Co_3O_4$ nanocomposite owning to presence of large amounts of interfaces in this sensor [21].

### 3.6 Selectivity studies

The gas selectivity properties of the $Fe_2O_3/Co_3O_4$ composite sensor were examined under the optimum condition. Some possible coexistence substances such as methanol, benzene and toluene, usually had an interference on the determination of ethanol gas in the traditional semiconductor oxide sensors, which would seriously limit the extensive utilization. To explore the selectivity of the $Fe_2O_3/Co_3O_4$ sensor for ethanol gas, the responses to the above mentioned gases were examined. The result is shown in Fig. 9. The present sensor showed a significantly high selectivity to ethanol gas. Different gases have different activation energies for adsorption, desorption and reaction on the metal oxide surface. Therefore, the response of the sensor would strongly depend on the gas being sensed at different temperatures. For the $Fe_2O_3/Co_3O_4$ nanocomposite sensor, 300°C is the optimal working temperature because the adsorption energy for ethanol is low at this temperature, whereas the activation energy for the adsorption of other gas species is relatively high at the temperature.

Table 1 summarizes ethanol sensing properties of some $Fe_2O_3$ based or $Co_3O_4$ based sensors with $Fe_2O_3/Co_3O_4$ nanocomposites sensor. The table shows that the present sensor has good sensitivity for ethanol detection ($R_a/R_g$ =10.86, 100 ppm) and especially it has a very short response time in comparison with other $Fe_2O_3$-based gas sensors.

## IV. CONCLUSION

In brief, $Fe_2O_3$ nanoparticles and $Fe_2O_3/Co_3O_4$ nanocomposites were successfully synthesized by a hydrothermal process. XRD studies showed high purity and good crystallinity of synthesized powders and SEM micrograph revealed good intimate mixing of synthesized $Fe_2O_3/Co_3O_4$ nanocomposites, indicating the effectiveness of hydrothermal method. The ethanol gas sensing performance of the synthesized powders were examined at different ethanol concentrations and temperatures. The $Fe_2O_3/Co_3O_4$ nanocomposites sensor showed superior sensing performance (R=10.86, $\tau_{res}$=1.36s, $\tau_{rec}$=40.25s for 100 pm ethanol at 300°C) to the pristine $Fe_2O_3$ sensor (R=4.44, $\tau_{res}$=1.56s, $\tau_{rec}$=41. 8s for 100 pm ethanol at 300°C). The improved gas sensing properties of the composite sensor were due mainly to the enhanced modulation of the conduction channel width and the enhanced modulation of the potential barrier formed at the $Fe_2O_3/Co_3O_4$ interface, accompanying the adsorption and desorption of ethanol gas, the stronger oxygen adsorption of p-type $Co_3O_4$, and the creation of preferential adsorption sites.

### ACKNOWLEDGMENTS

This research was supported by the MSIP(Ministry of Science, ICT and Future Planning), Korea, under the C-ITRC(Convergence Information Technology Research Center) (IITP-2015-H8601-15-1003) supervised by the IITP(Institute for Information & communications Technology Promotion and Basic Science Research Program through the National Research Foundation of Korea (NRF) funded by the Ministry of Education (2015-0020163).

**Table 1.** Comparison ethanol sensing characteristics of some $Fe_2O_3$ based sensors reported in the literature with the present work.

| Material | Conc. (ppm) | Response (Ra/Rg) | Temp. (°C) | $\tau_{res}/\tau_{rec}$ (sec) | Ref. |
|---|---|---|---|---|---|
| Pristine α-$Fe_2O_3$ NPs | 100 | $68^c$ | 325 | 50 | [22] |
| Ag@$Fe_2O_3$ | 100 | $6.3^a$ | 250 | 5.5/16 | [12] |
| $Fe_2O_3/SnO_2$ | 100 | $230^c$ | 235 | -/- | [23] |
| $Fe_2O_3$ /$TiO_2$ | 100 | $35^b$ | 240 | -/- | [23] |
| $Fe_2O_3$/CdO | 100 | $20^b$ | 300 | ~6/~10 | [24] |
| $Fe_2O_3$/ZnO nanocomposite | 10 | $4.7^b$ | 220 | ~20/~20 | [25] |
| α-$Fe_2O_3$ | 200 | $1.3^b$ | 200 | -/- | [26] |
| $Fe_2O_3$/ZnO nanorods | 100 | $7.34^b$ | 200 | -/- | [27] |
| $Fe_2O_3/Co_3O_4$ nanocomposite | 100 | $10.86^b$ | 300 | 1.36/40.25 | Present Work |

a:($R_a/R_g$, %), b:($R_a/R_g$), c:($[R_a-R_g]/R_a$, %),

## Figure Captions.

**Fig. 1.** SEM images of (a) pure $Fe_2O_3$ nanoparticles, (b) pure $Co_3O_4$nanoparticles, (c) $Fe_2O_3/Co_3O_4$ composite nanoparticles, (d) EDX elemental map, and (e) EDX spectrum of $Fe_2O_3/Co_3O_4$ nanocomposite.

**Fig. 2**. XRD patterns of $Fe_2O_3$ nanoparticles and $Fe_2O_3$/ $Co_3O_4$ nanocomposites.

**Fig. 3.** (a) Low-magnification TEM image of $Fe_2O_3$/ $Co_3O_4$ nanocomposites. (b) High Resolution-TEM image of the interfacial region of $Fe_2O_3$/ $Co_3O_4$ nanocomposites. (c) Corresponding selected area diffraction pattern.

**Fig. 4.** Response of $Fe_2O_3$ and $Fe_2O_3$ and $Co_3O_4$, sensors towards 200 ppm ethanol vapor at different temperatures.

**Fig. 5.** Dynamic response of (a) $Fe_2O_3$ NPs sensors and (b) $Fe_2O_3/Co_3O_4$ nanocomposite sensors towards 5, 10, 20, 50, 100 and 200 ppm ethanol vapor at 300°C.

**Fig. 6.** (a) Calibration curve of $Fe_2O_3$ and $Fe_2O_3/Co_3O_4$ sensors at 300°C. (b) Logarithmic plot of response of sensors towards different ethanol concentration at 300°C.

**Fig. 7.** Schematic diagrams and corresponding energy band diagrams showing the depletion layers and potential barrier height formed in (a) an $Fe_2O_3$-$Fe_2O_3$ nanoparticle couple abundant in pure $Fe_2O_3$ and (b) an $Fe_2O_3/Co_3O_4$ nanoparticle couple abundant in $Fe_2O_3/Co_3O_4$ nanocomposites.

**Fig. 8.** Plot of response (a) and recovery (b) times for $Fe_2O_3$ and $Fe_2O_3/Co_3O_4$ sensors towards different concentrations of ethanol vapor at 300°C.

**Fig. 9.** Selectivity pattern of $Fe_2O_3/Co_3O_4$ nanocomposite sensor at 300°C.

**Figures**

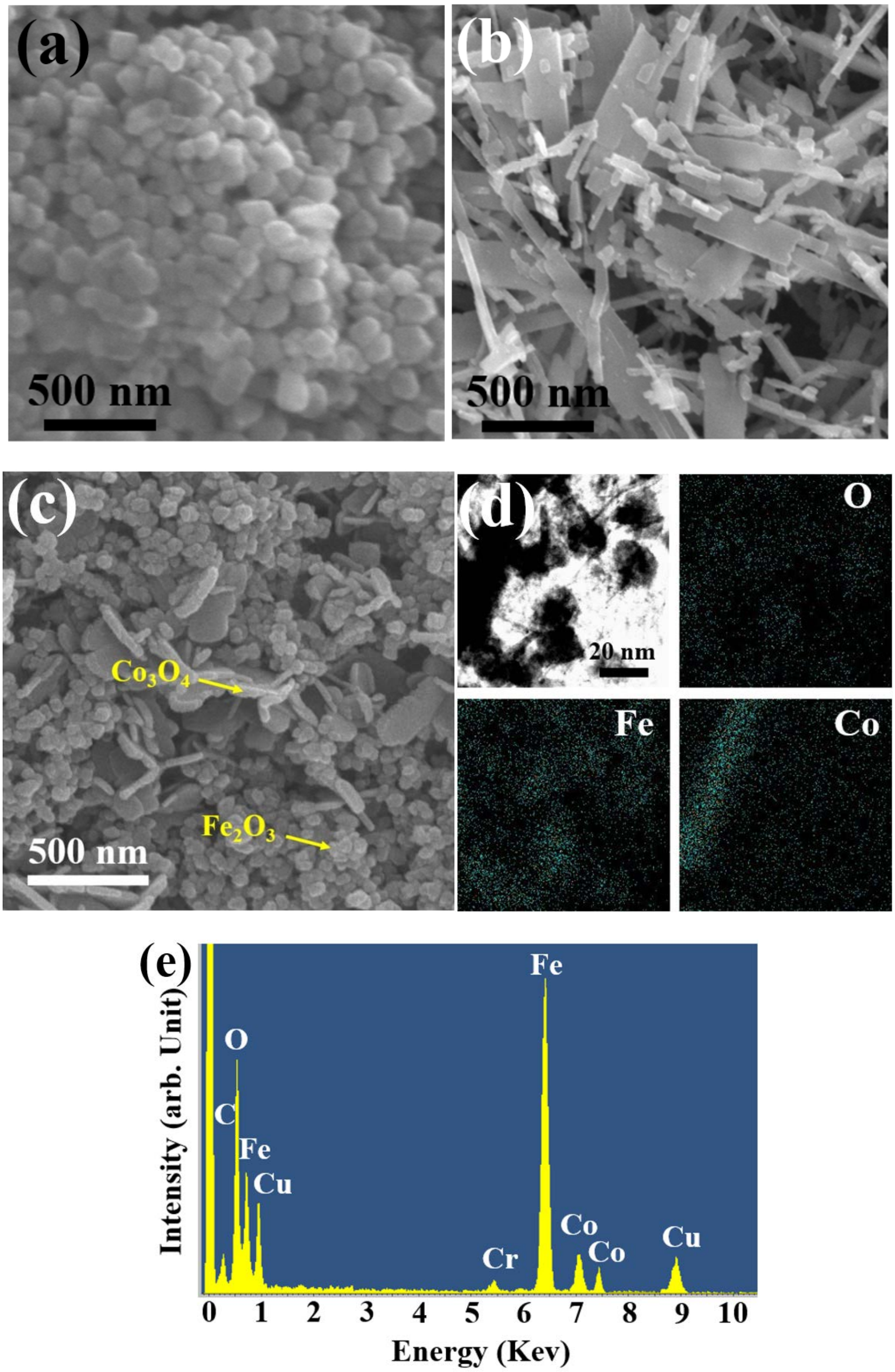

**Fig. 1.**

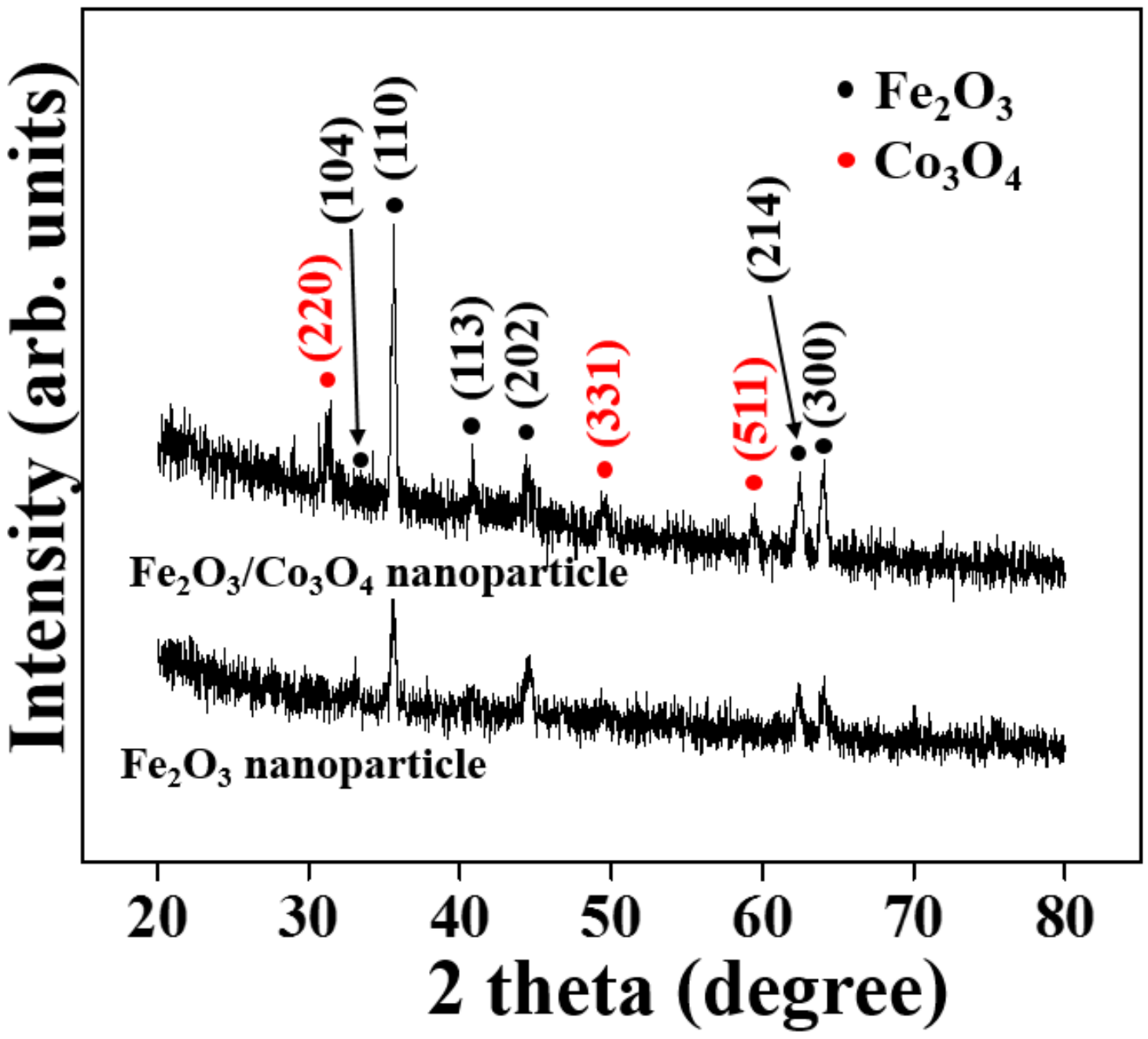

**Fig. 2.**

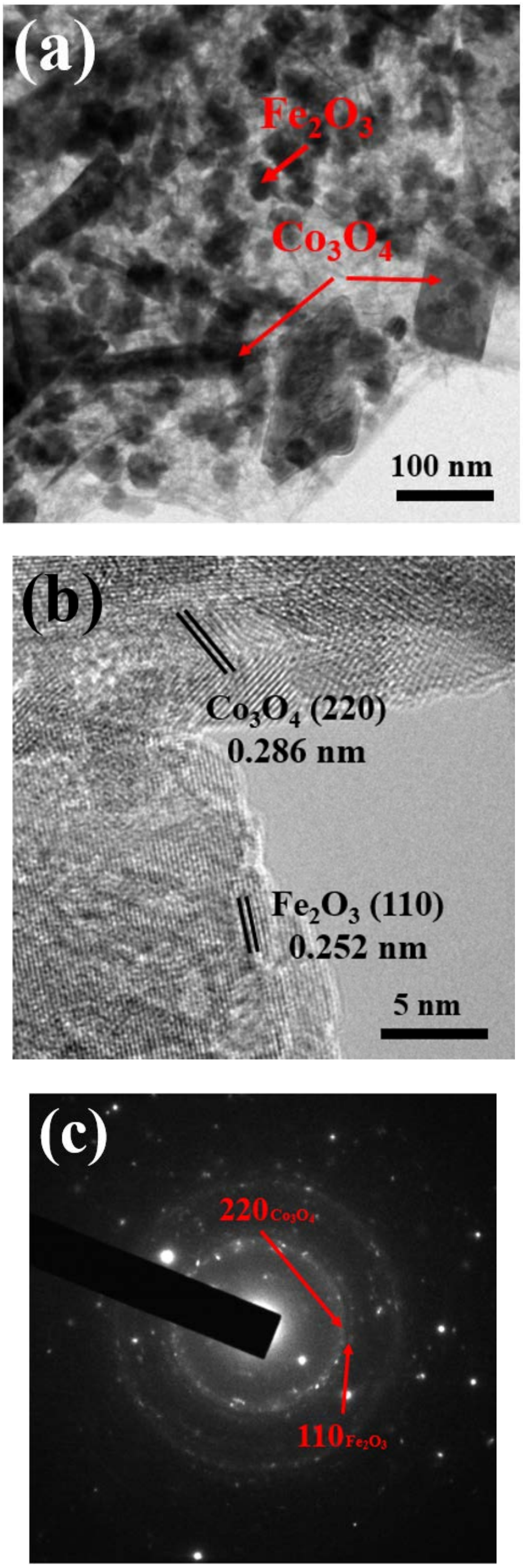

**Fig. 3.**

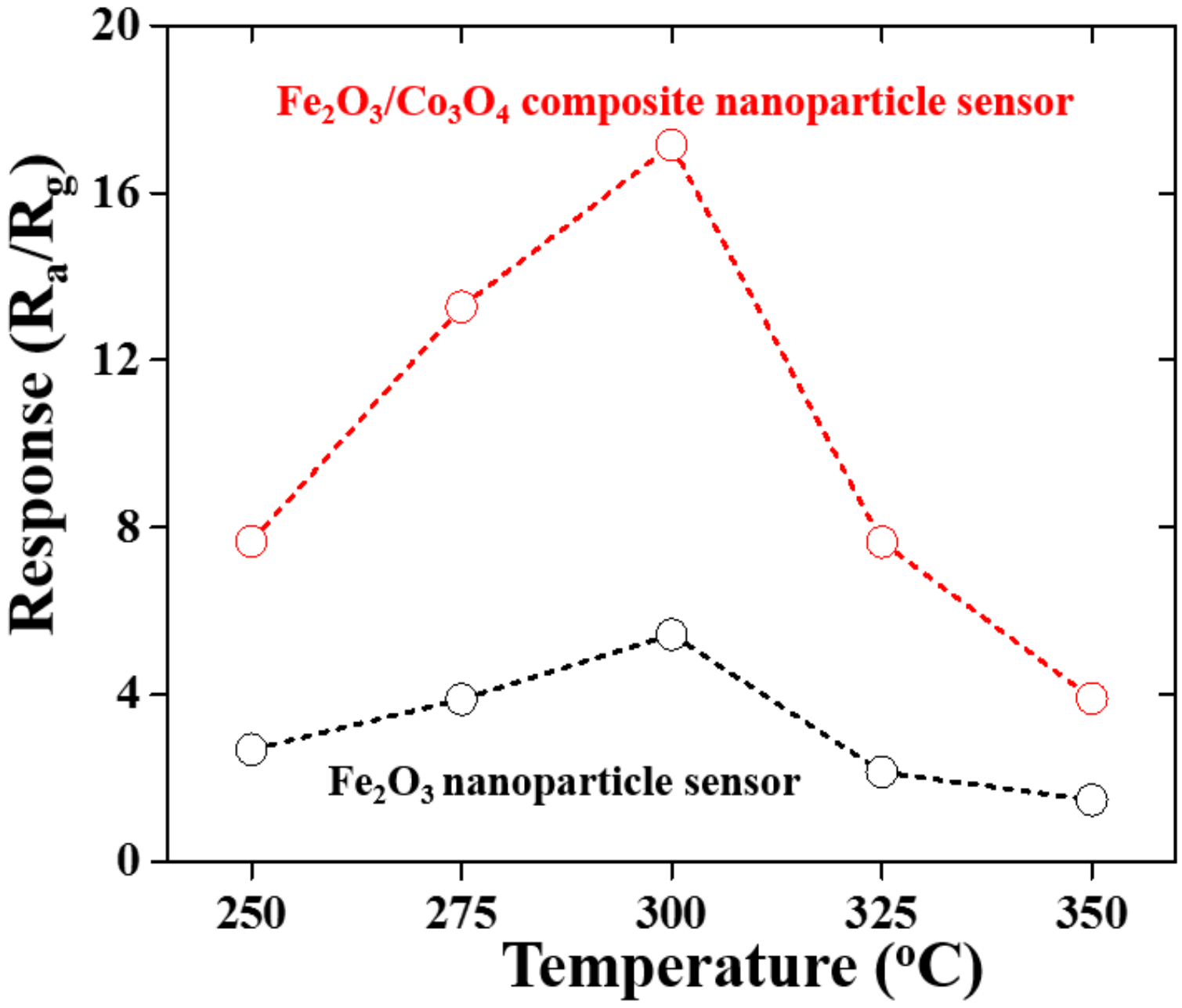

**Fig. 4.**

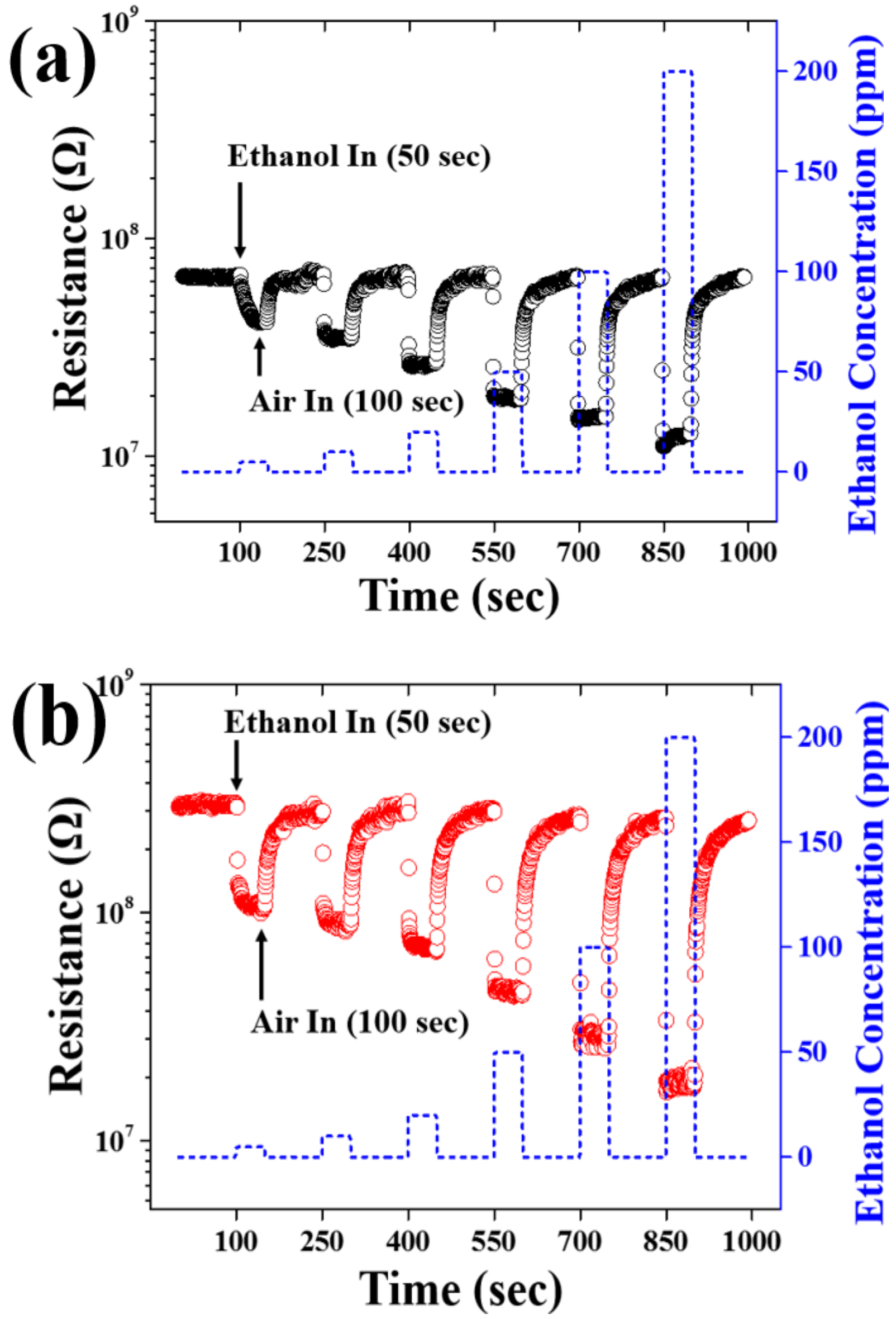
(a)
Ethanol In (50 sec)
Air In (100 sec)
Resistance (Ω)
Ethanol Concentration (ppm)
Time (sec)
(b)
Ethanol In (50 sec)
Air In (100 sec)
Resistance (Ω)
Ethanol Concentration (ppm)
Time (sec)

**Fig. 5.**

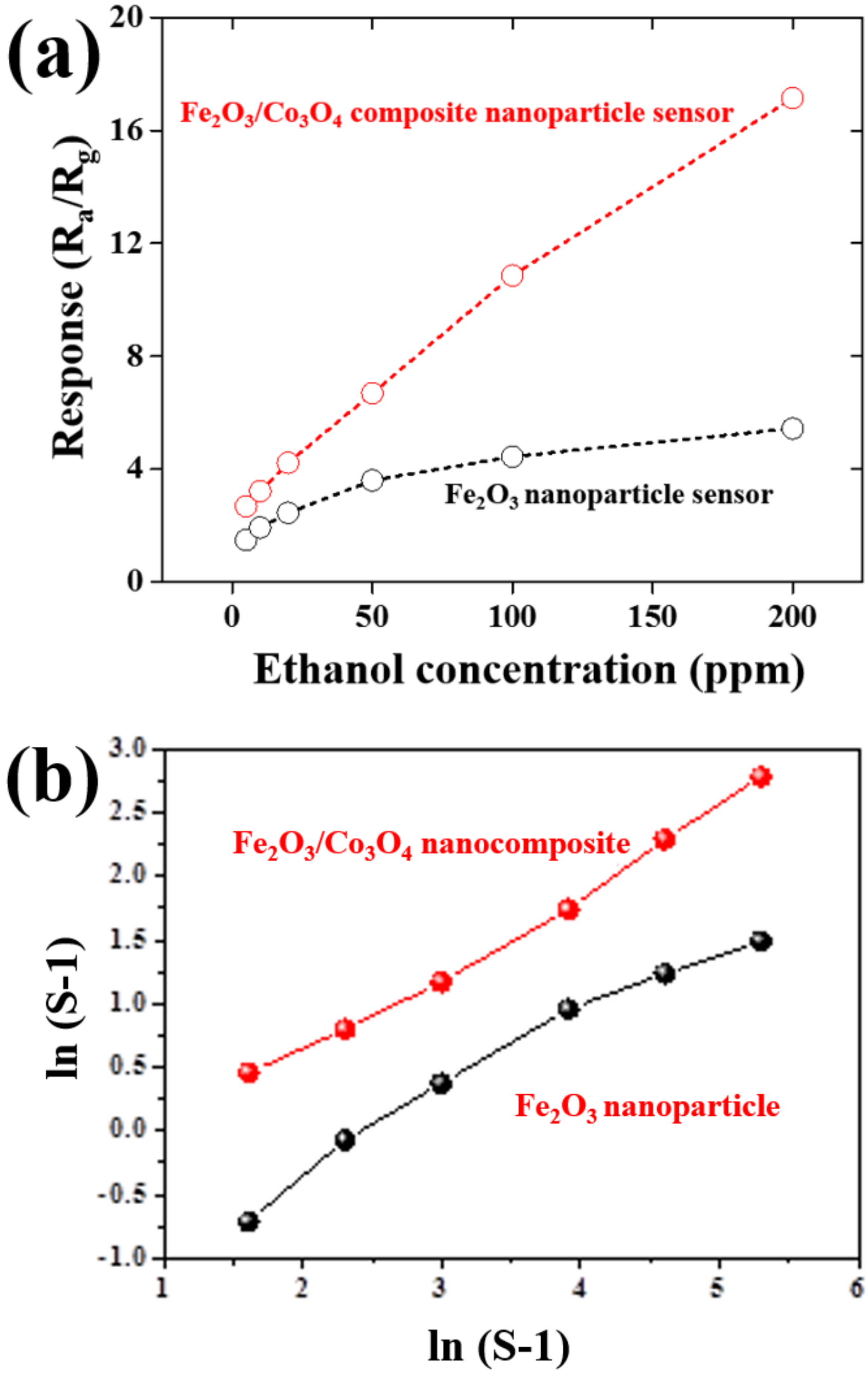

**Fig. 6.**

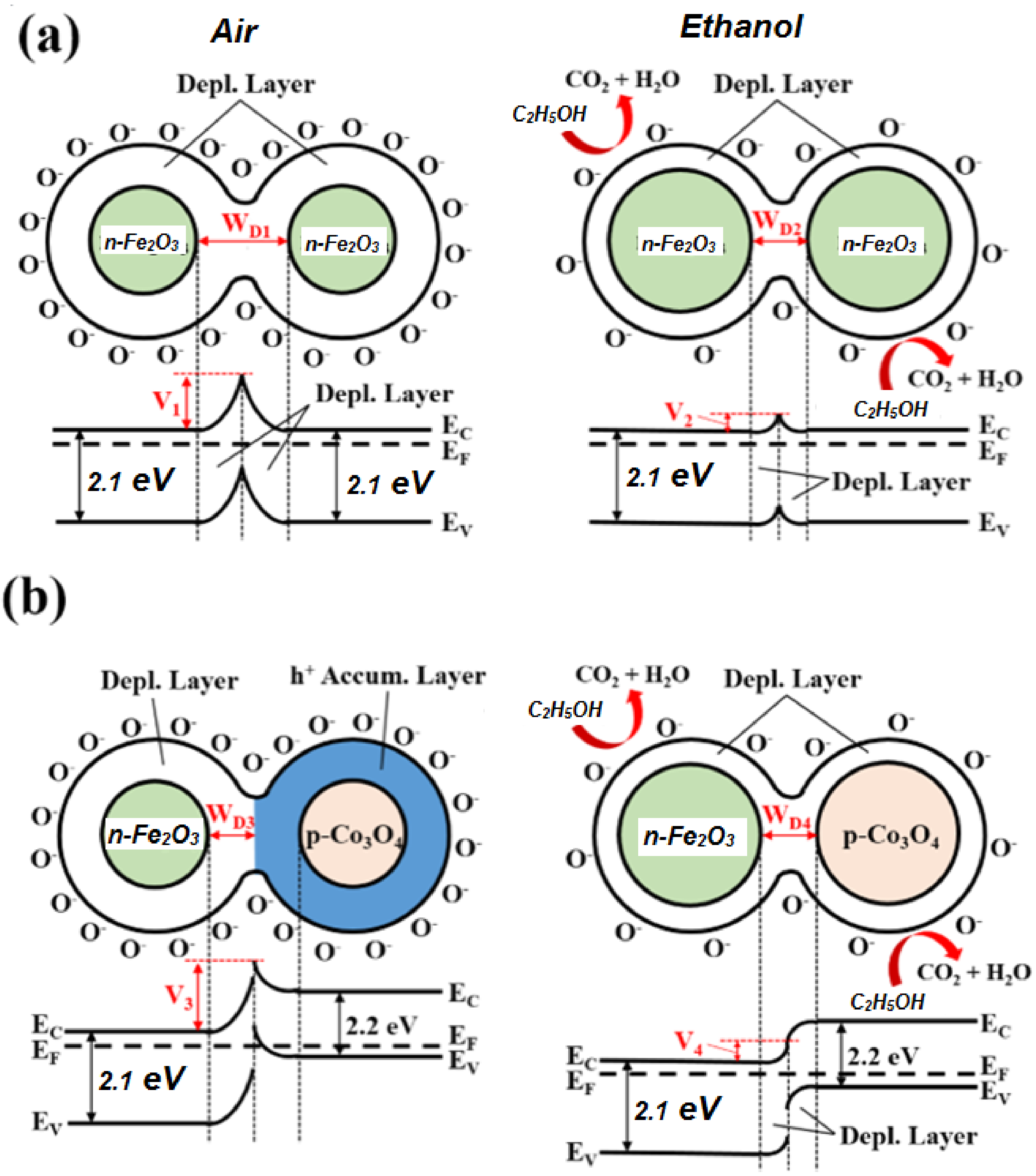

**Fig. 7.**

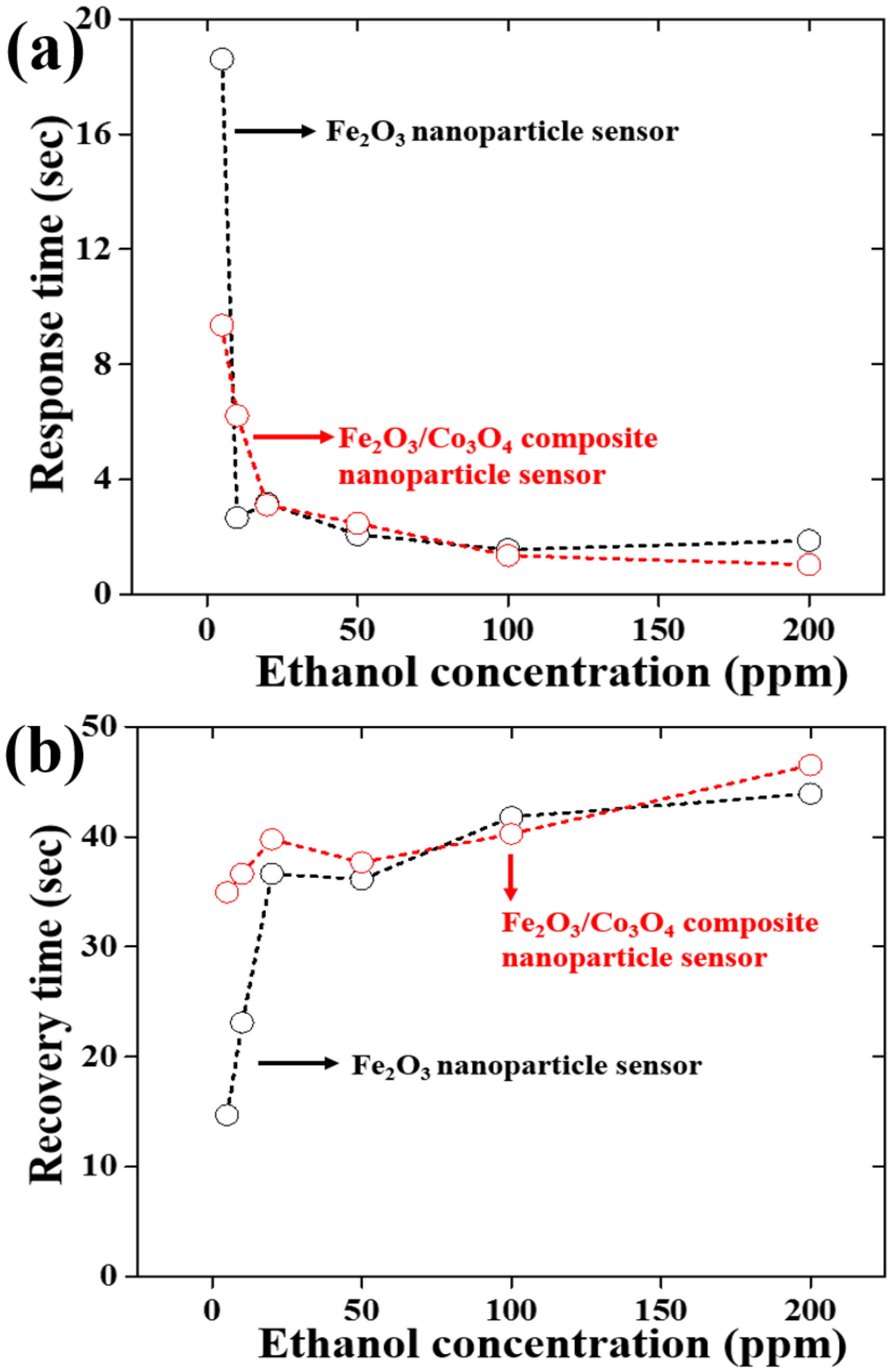

**Fig. 8.**

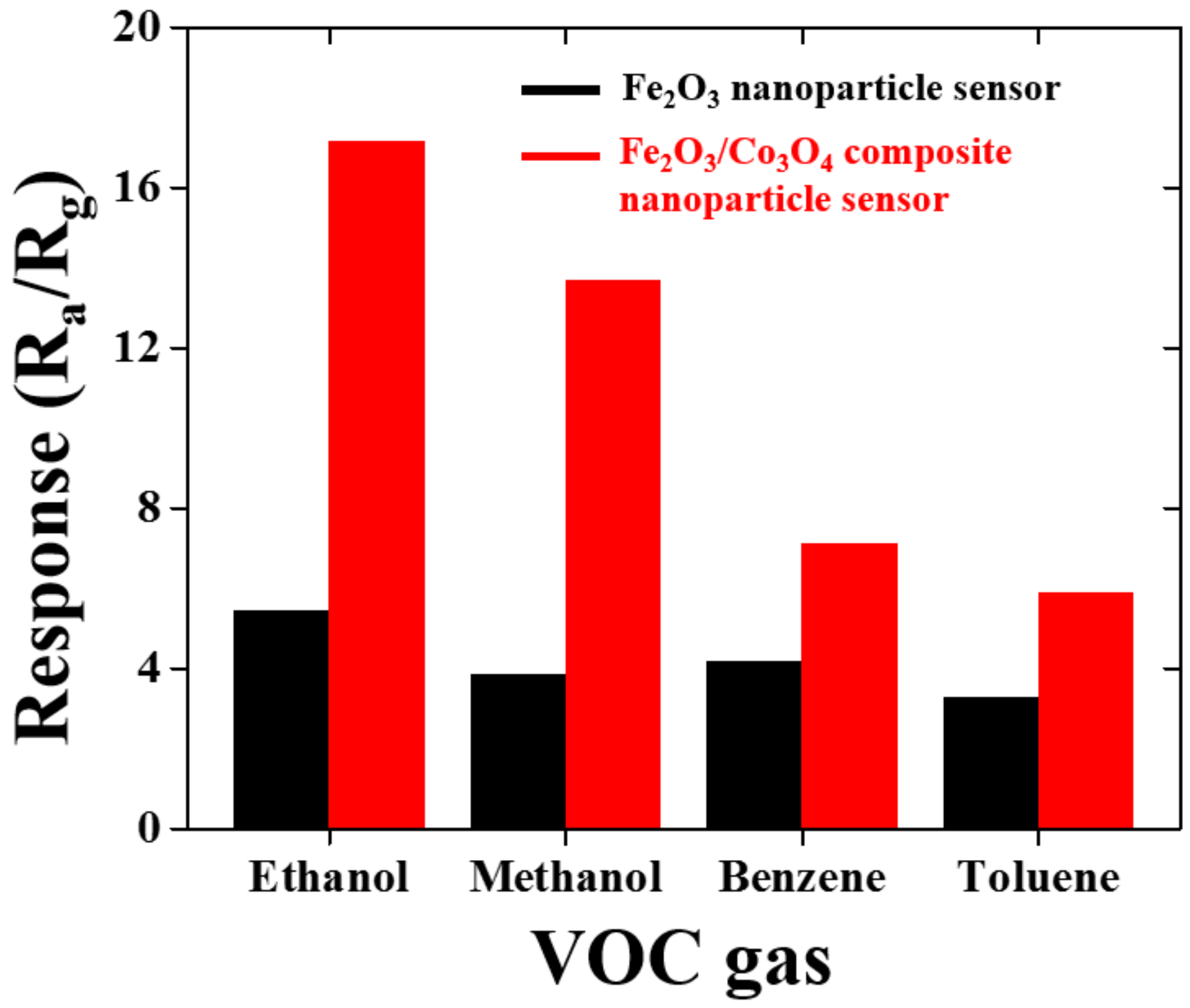

**Fig. 9.**